\documentclass[twocolumn,amssymb,amsfonts,floatfix]{revtex4}
\usepackage{bm}
\usepackage{dcolumn}
\usepackage{mathbbol}
\usepackage{graphics,graphicx}
\usepackage{epstopdf}
\usepackage{amsmath,amsthm}
\usepackage{mathrsfs}
\usepackage{times}
\usepackage{color}
\usepackage{calc}
\usepackage{epsfig}
\usepackage{float}

\begin{document}
\title{ Excited State Quantum Phase Transitions Studied from a Non-Hermitian Perspective }
\author{Milan \v{S}indelka\footnote{sindelka@fzu.cz}}
\affiliation{Institute of Plasma Physics, Academy of Sciences of the Czech Republic, Prague 8, 18200, Czech Republic} 
\author{Lea F. Santos\footnote{lsantos2@yu.edu}}
\affiliation{Department of Physics, Yeshiva University, New York, New York 10016, USA}
\author{Nimrod Moiseyev\footnote{nimrod@technion.ac.il}}
\affiliation{Schulich Faculty of Chemistry, Institute of Solid State, and Faculty of Physics, Technion - Israel Institute of Technology,
 Haifa 32000 Israel}
\date{\today}
\begin{abstract}
A main distinguishing feature of non-Hermitian quantum mechanics is the presence of exceptional points (EPs). They correspond to the coalescence of two energy levels and their respective eigenvectors. Here, we use the Lipkin-Meshkov-Glick (LMG) model as a testbed to explore the strong connection between EPs and the onset of excited state quantum phase transitions (ESQPTs). We show that for finite systems, the exact degeneracies (EPs) obtained with the non-Hermitian LMG Hamiltonian continued into the complex plane are directly linked with the avoided crossings that characterize the ESQPTs for the real (physical) LMG Hamiltonian. The values of the complex control parameter $\alpha$ that lead to the EPs approach the real axis as the system size $N\rightarrow \infty$. This happens for both, the EPs that are close to the separatrix that marks the ESQPT and also for those that are far away, although in the latter case, the rate the imaginary part of $\alpha$ reduces to zero as $N$ increases is smaller. With the method of Pad\'e approximants, we can extract the critical value of $\alpha$. 
\end{abstract}
\maketitle

{\em Introduction.--}
A quantum phase transition (QPT) corresponds to the vanishing of the gap between the ground state and the first excited state in the thermodynamic limit~\cite{CarrBook,SachdevBook}. Excited state quantum phase transitions (ESQPTs) are generalizations of QPTs to the excited levels~\cite{Cejnar2006,Caprio2008}. They emerge when the QPT is accompanied by the bunching of the eigenvalues around the ground state. This divergence in the density states at the lowest energy moves to higher energies as the control parameter increases above the QPT critical point. The energy value where the density of states peaks marks the point of the ESQPT.

ESQPTs have been analyzed in various theoretical models~\cite{Caprio2008,Bernal2008,Ribeiro2008,Relano2008,Cejnar2009,Fernandez2009,Bernal2010,Fernandez2011,Fernandez2011b,Yuan2012,Brandes2013,Bastarrachea2014b,Stransky2015,Engelhardt2015,SantosBernal2015,Bernal2016,Santos2016,FernandezARXIV} and have also been observed experimentally~\cite{Winnewisser2005, Zobov2006,Larese2011,Larese2013,Dietz2013,Zhao2014}. They have been linked with the bifurcation phenomenon~\cite{Santos2016} and with the exceedingly slow evolution of initial states with energy close to the ESQPT critical point~\cite{SantosBernal2015,Bernal2016,Santos2016}. Equivalently to what one encounters in QPTs, the nonanalycities associated with ESQPTs occur in the thermodynamic limit. When dealing with finite systems, signatures of these transitions are usually inferred from scaling analysis.  There are, however, studies based on new microcanonical distributions that claim that QPTs can be predicted without considerations of thermodynamic limits~\cite{GrossBook,Brody2007}. One might expect analogous results for ESQPTs.

In this work, we show that the nonanalycities associated with QPTs and ESQPTs can be found in finite systems when the control parameter of the Hamiltonian is continued into the complex plane. The Hamiltonian that we study,
\begin{equation}
H(\alpha,N)=\alpha H_I(N) +(1-\alpha) H_{II}(N),
\label{ham_example}
\end{equation}
is a linear combination of two noncommuting operators, $[H_I,H_{II}]\ne {0}$, where $\alpha$ is the control parameter and $N$ is the system size. As the control parameter varies from $\alpha=1$ to $\alpha=0$ the spectrum of the full Hamiltonian is transformed from the spectrum of $ H_I$ to the spectrum of $ H_{II}$. The transition of the ground and excited states from one symmetry to a mixture of different symmetry solutions is continuous in $\alpha$, yet quite sharp. It is only in the limit of $N\to \infty$ that a point of nonanalyticity appears for the ground state at a critical value $\alpha_c$ and for the excited states at values of $\alpha_{\text{ESQPT}} <\alpha_c$. In finite systems, the sharp transition from one type of symmetry adapted solution to another one is associated with avoided crossings. We show that these avoided crossings are connected with the exceptional points (EPs) of the non-Hermitian form of $ H(\alpha,N)$, where $\alpha$ is complex. 

The association between EPs and avoided crossings was first presented in~\cite{Heiss1990}. Connections have also been made between EPs and QPTs~\cite{Heiss1988,Heiss2002,Cejnar2007PRL,Heiss2012} and between EPs and ESQPTs~\cite{Heiss2005}. Here, we further elaborate the studies of ESQPTs from the perspective of non-Hermitian Hamiltonians taking into account both EPs close and also far apart from the real axis.

The EPs that we calculate correspond to the exact degeneracies of the non-Hermitian Hamiltonian found for specific values $\alpha_{\text{EP}}$ of the complex control parameter. More precisely, they correspond to branch point singularities of the eigenvalues and eigenvectors~\cite{Moiseyev1980,MoiseyevBook,KatoBook,Heiss2004,Berry2004}. We show that as $N\rightarrow \infty$, the complex  $\alpha_{\text{EP}}$ approach and accumulate at the real axis, therefore coinciding with the QPT and ESQPT critical values of the real (physical) Hamiltonian. We notice that all EPs approach the critical values, those close to the separatrix that marks the ESQPT and also those far away. However, the distant ones converge to those values more slowly. Using the Pad\'e extrapolation technique, we demonstrate that these critical values can be derived from the EPs obtained with finite system sizes.

{\em Model and separatrix.--}
ESQPTs have been extensively studied in Hamiltonians with a $U(n+1)$ algebraic structure given by $H_{U(n+1)} = \alpha H_{U(n)}  - (1-\alpha)N^{-1} H_{SO(n+1)}$. They are composed of two limiting dynamical symmetries, the $U(n)$ and the $SO(n+1)$. In the bosonic form, these Hamiltonians represent limits of the vibron model~\cite{Iachello1981,IachelloBook,Iachello1996,Bernal2005}, which is used to characterize the vibrational spectra of molecules. The $U(n)$ dynamical symmetry ($\alpha=1$) is described by a one-body operator and the $SO(n+1)$ dynamical symmetry ($\alpha=0$) by a two-body operator, so the latter needs to be rescaled by the system size $N$.

These $U(n+1)$ Hamiltonians show a second-order ground state QPT at $\alpha_c =0.8$  and ESQPTs for $\alpha_{\text{ESQPT}} <\alpha_c$. Our analysis is illustrated for the $U(2)$ Hamiltonian, which represents one of the spin versions of the LMG model~\cite{Lipkin1965a,Cejnar2009}. The Hamiltonian is written as~\cite{Yuan2012,Santos2016},
\begin{equation}
H_{U(2)} =  \alpha \left( \frac{N}{2} + {\cal S}_z \right) -  \frac{4(1-\alpha)}{N} {\cal S}_x^2 ,
\label{Hbosons}
\end{equation}
where ${\cal S}_z = \sum_{i=1}^{N} S_i^z$ is the total spin in the $z$-direction and ${\cal S}_x = \sum_{i=1}^{N} S_i^x$ is   the total spin in the $x$-direction. The first term favors the alignment of the spins in the $z$ direction and the second term in the $x$ direction.

\begin{figure}[htb]
\centering
\includegraphics*[width=3.2in]{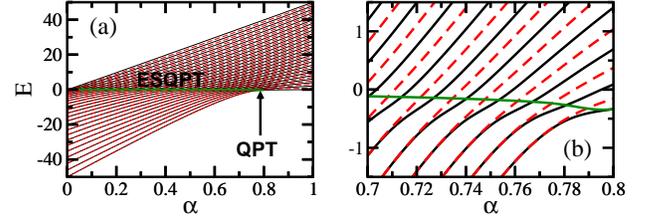}
\caption{(Color online) Energy levels \textit{vs} $\alpha$ for $N = 50$ (a) and $N=100$ (b). Eigenstates of one parity are indicated with black solid lines and from the other with dashed red lines. The horizontal green line is the separatrix, it indicates $E_c$ obtained from  Eq.~(\ref{Eq:separatrix}). Arbitrary units.}
\label{Fig:Eigenvalues}
\end{figure}

For $\alpha=1$, all eigenvalues of $H_{U(2)}$ are positive. For $\alpha=0$, the eigenvalues  are negative and the eigenstates form pairs of degenerate states, one with positive and the other with negative total magnetization in $x$. In Fig.~\ref{Fig:Eigenvalues} (a), we show the eigenvalues versus the control parameter for $N=50$.  The ground state QPT occurs at $\alpha_c =0.8$ for $E_c \simeq 0$. For $\alpha <\alpha_c$, the lowest energies become smaller than zero, while the states with energy close to $E_c$ cluster together. The bunching of the energy levels at $E_c $ characterizes the ESQPT.

The solid (nearly) horizontal line in Figs.~\ref{Fig:Eigenvalues} (a) and (b) is the separatrix that marks the ESQPT. It can be obtained from a semiclassical analysis. The normalized energy difference between $E_c$ and the ground state eigenvalue $E_{\text{GS}}$ is the critical excitation energy of the ESQPT. Its equation is given by~\cite{Caprio2008,Bernal2008,Bernal2010}
\begin{equation}
E_{\text{ESQPT}} (\alpha) = \frac{E_c-E_{\text{GS}}}{N}=\frac{[1 - 5(1-\alpha)]^2 }{16 (1-\alpha) }.
\label{Eq:separatrix}
\end{equation}
In Fig.~\ref{Fig:Eigenvalues}, the line for the separatrix corresponds to the value of $E_c$ obtained using $E_{\text{ESQPT}} $ from Eq.~(\ref{Eq:separatrix}) and the numerical data for $E_{\text{GS}}$. 

In Fig.~\ref{Fig:Eigenvalues} (b), we consider $N=100$ and zoom in the data for $0.7 \leq \alpha \leq 0.8$. This figure makes clear the effect of the phase transition on the structure of the eigenstates. The eigenstates with energy $E<E_c$ are almost doubly degenerate. These are the states with structures closer to the $SO(2)$ symmetry. The degeneracy is lifted for $E>E_c$, where the eigenstates have structure closer to the $U(1)$ symmetry. Quantities such as the participation ratio~\cite{SantosBernal2015,Santos2016} and the fidelity~\cite{Bernal2016} have been used to capture the abrupt changes in the structures of eigenstates caused by ESQPTs.

{\em Non-Hermitian Formalism and EP.--}
In the vicinity of a critical point of a finite system described by a Hermitian Hamiltonian $H(\alpha,N)$, the crossings of the energy levels are avoided. In contrast, the complex eigenvalues of the corresponding non-Hermitian Hamiltonian, obtained by continuing the control parameter $\alpha$ into the complex plane, can cross. This degeneracy, accompanied by the coalescence of the correspondent eigenvectors, is the EP.  We find various EPs for different complex values $\alpha_{\text{EP}}(N)$ of the control parameter. It has in fact been proven that in the case of a Hamiltonian such as that in Eq.~(\ref{ham_example}), where the two Hermitian operators $H_I$ and $H_{II}$ do not commute, there always exists a complex linear pre-factor for which the EPs are obtained~\cite{Moiseyev1980}.  Our results below substantiate the strong relationship between critical points and the appearance of EPs.

Sufficiently close to $\alpha_{\text{EP}}(N)$ the energy spectrum of the non-Hermitian Hamiltonian contains two almost degenerate values given by
\begin{equation}
E_\pm(\alpha,N)\cong E_{\text{EP}}(\alpha_{\text{EP}} (N))\pm C(N)\sqrt{\alpha-\alpha_{\text{EP}} (N)},
\label{EdegNonHerm}
\end{equation}
where $C(N)$ is a function of the system size. The two eigenvectors corresponding to these energies are
\begin{equation}
|{\bf \psi}_\pm(\alpha, N)\rangle\cong |{\bf \psi}_{\text{EP}}(N) \rangle \pm |{\bf \chi}(N)\rangle\sqrt{\alpha-\alpha_{\text{EP}} (N)} .
\end{equation}
The orthogonality condition, which can be extended to symmetric non-Hermitian Hamiltonians, implies that the inner product $\langle {\bf \psi}^*_\mp(N)|{\bf \psi}_\pm(N) \rangle =0$.
At the critical complex value $\alpha_{\text{EP}}$ of the control parameter, the degenerate states become self-orthogonal, that is~\cite{MoiseyevBook}
\begin{eqnarray}
&&|{\bf \psi}_{+} (\alpha_{\text{EP}}, N)\rangle = |{\bf \psi}_{-} (\alpha_{\text{EP}}, N)\rangle = |{\bf \psi}_{\text{EP}} (N)\rangle , \label{eq:cusp}  \\
&&\langle {\bf \psi}^*_{\text{EP}}(N)|{\bf \psi}_{\text{EP}}(N)\rangle=0. \nonumber
\end{eqnarray}
Because of the self-orthogonality at $\alpha_{\text{EP}}$, the quantum fluctuations at this point become infinitely large if associated with the expectation value of  $\partial H(\alpha,N)/\partial \alpha$. This gives further support to associating QPT and ESQPT with EPs.

\begin{figure}[ht]
\vskip -0.4 cm
\includegraphics*[width=2.5in]{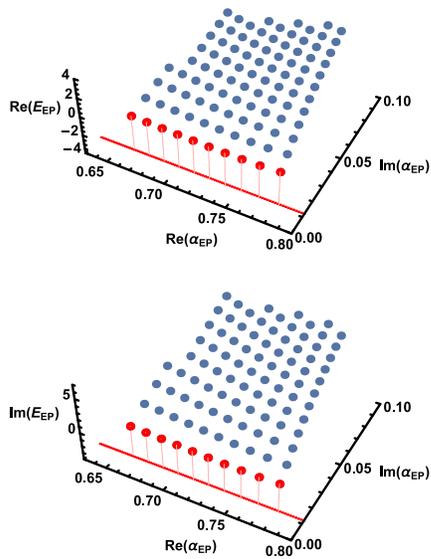}
\caption{ Exceptional points (circles) of the complex dilated LMG Hamiltonian (\ref{Hbosons})  for $N=100$. Top panel: the real part of $E_{\rm EP}$, and bottom panel: the imaginary part of $E_{\rm EP}$. The  EPs with the lowest ${\rm Im}(\alpha_{\rm EP})$ are indicated with a light (red) color. They have almost real valued energies and ${\rm Im}(\alpha_{\rm EP}) \sim 0.0115$; the latter is shown with a solid line on the $\alpha$ plane.}
\label{Fig:EPs}
\end{figure}

In Fig.~\ref{Fig:EPs}, we use circles to represent the EPs of the complex LMG Hamiltonian, which is obtained from  Eq.~(\ref{Hbosons}) by continuing $\alpha$ in the complex plane. The real part of the energies $E_{\text{EP}}$ are shown in the top panel and the imaginary part in the bottom panel. The EPs with the lowest imaginary part of $\alpha_{\text{EP}}$ are indicated with a light (red) color. They have ${\rm Im}(\alpha_{\rm EP}) $ almost constant and close to 0.0115 for $N=100$. This array of EPs also has the lowest  ${\rm Im}(E_{\text{EP}})$. For higher ${\rm Im}(E_{\text{EP}})$, we find other rows of EPs also with approximately constant values of ${\rm Im}(\alpha_{\rm EP}) $.

\begin{figure}[ht]
\includegraphics*[width=3.2in]{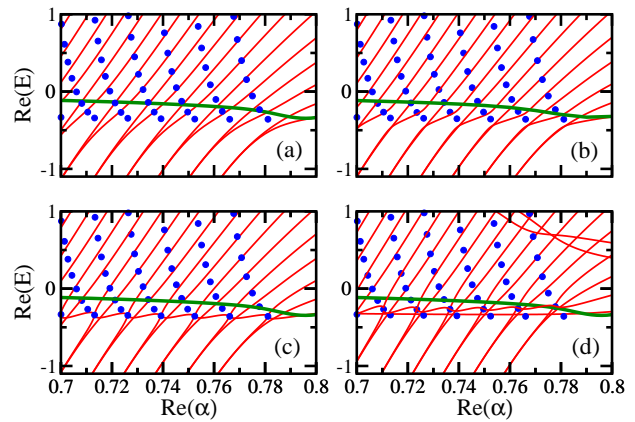}
\caption{Real part of the eigenvalues of the complex dilated LMG Hamiltonian (\ref{Hbosons})  {\em vs.}  the real part of the complex $\alpha$ (thin solid lines) for $N=100$ and ${\rm Im}(\alpha) = 0$ (a), $0.0900$ (b),  $0.0115$ (c), $0.0230$ (d). Circles are the EPs, they correspond to ${\rm Re}(E_{\text{EP}})$ versus ${\rm Re}(\alpha_{\text{EP}})$. The thick nearly horizontal line is the separatrix.}
\label{Fig:esptEP}
\end{figure}

{\em ESQPT vs. EP.--} To unveil the connection between ESQPTs and EPs, we now compare the results from the Hermitian and non-Hermitian approaches. In Fig.~\ref{Fig:esptEP}, the thin lines give the real part of the eigenvalues of the complex LMG Hamiltonian as a function of ${\rm Re}(\alpha)$, the circles are the EPs as in Fig.~\ref{Fig:EPs}, and the thick nearly horizontal line is the separatrix. In each panel, ${\rm Re}(\alpha)$ varies from 0.7 to 0.8, while ${\rm Im}(\alpha)$ is held at a constant value.

In Fig.~\ref{Fig:esptEP} (a), ${\rm Im}(\alpha)=0$, so the plot is the same as in Fig.~\ref{Fig:Eigenvalues} (b), but now with the EPs added to it. This figure already suggests a strong link between the EPs and the ESQPT. As one sees, for $E<E_c$, where we have pairs of degenerate states, there are no EPs. As the energies increase, they first appear very close to the point where the degeneracy is lifted and in the vicinity of the separatrix.

To better support this relationship, we increase the value of ${\rm Im}(\alpha)$ from Fig.~\ref{Fig:esptEP} (a) to (c) up to ${\rm Im}(\alpha)=0.0115$. The latter is the value of the sequence of EPs with the lowest ${\rm Im}(\alpha_{\text{EP}})$, as shown in Fig.~\ref{Fig:EPs}. By increasing ${\rm Im}(\alpha)$, the values of ${\rm Re}(E)$ of the non-Hermitian Hamiltonian change, while the EPs and the separatrix naturally remain the same. The thin solid lines are continuously deformed from Fig.~\ref{Fig:esptEP} (a) to (c) until the avoided crossings become true crossings. They happen right at the EPs with the lowest ${\rm Re}(E_{\text{EP}})$. These EPs are located on the bifurcating branches of the spectrum. Lines intersecting at the EPs exhibit cusps, which is consistent with Eq.~(\ref{EdegNonHerm}). These observations indicate that the ESQPT in Fig.~\ref{Fig:Eigenvalues} is inherently caused by the non-Hermitian crossings (cusps). We can thus interpret ESQPTs as phenomena arising due to the presence of EPs.

In  Fig.~\ref{Fig:esptEP} (d), we choose ${\rm Im}(\alpha)=0.0230$, which is close to the value for the second row of EPs. Analogously to Fig.~\ref{Fig:esptEP} (c), there are true crossings (cusps) coinciding with the locations of these EPs. As ${\rm Im}(\alpha)$ further increases, the crossings happen for sequences of EPs with higher and higher ${\rm Re}(E)$.
We show next that as $N$ increases, one by one, these sequences of EPs approach and accumulate on the real axis. Close to the critical points, there is a high density of EPs.

\begin{figure}[h!]
\includegraphics*[width=3.3in]{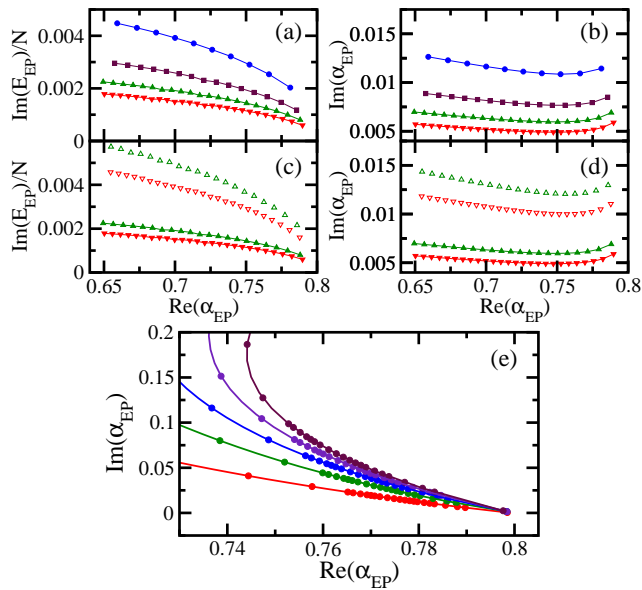}
\caption{EPs with the lowest  and the second lowest  values of ${\rm Re}(E_{\text{EP}})$ for different $N$'s (a)--(d). Extrapolation towards $N \to \infty$ carried out with the method of Pad\'{e} approximants (e). The lowest EPs in (a) and (b): $N=100$ (circles), $N=150$ (squares), and $N=200$  (up triangles), $N=250$ (down triangles). In (c) and (d): the lowest EPs (filled symbols) and the second lowest EPs (empty symbols) for $N=200$  (up triangles) and $N=250$ (down triangles).
In (e): the circles correspond to 22 system sizes between $N=20$ and $N=250$. Each line is a specific pair of states $(j_1,j_2)$ forming an EP, from bottom to top: $(0,2)$, $(1,3)$, $(1,4)$, $(1,5)$, $(1,6)$. The solid lines are obtained via the Pad\'e method, leading to the extrapolated points at ${\rm Im}(\alpha_{\text{EP}}) \rightarrow 0$, where ${\rm Re}(\alpha_{\text{EP}}) \rightarrow \alpha_c=0.8$.}
\label{PADE}
\end{figure}

{\em Thermodynamic limit.--} For a given system size, we have a discrete collection of EPs. As the system size increases, the number of EPs increases and they approach the separatrix (which in turn approaches zero, $E_c/N \rightarrow 0$).  This is illustrated in Figs.~\ref{PADE} (a) and (b) for the EPs with the lowest values of ${\rm Re}(E_{\text{EP}})$ and ${\rm Im}(E_{\text{EP}})$. As $N$ increases,  ${\rm Im}(E_{\text{EP}})/N$ and ${\rm Im}(\alpha_{\text{EP}})$ go to zero [the same occurs for ${\rm Re}(E_{\text{EP}})/N$ (not shown)] . In the thermodynamic limit, ${\rm Im}(\alpha_{\rm EP})  \rightarrow 0$, ${\rm Im}(E_{\text{EP}})/N \rightarrow 0$, ${\rm Re}(E_{\rm EP})/N \rightarrow E_c/N $, and ${\rm Re}(\alpha_{\rm EP}) $ coincides with $\alpha_{\text{ESQPT}}$.

In Figs.~\ref{PADE} (c) and (d), we compare the EPs with the lowest (filled symbols) and the second lowest  (empty symbols) energies for $N=200$ and $250$. The second lowest EPs also approach the separatrix as the system size increases, but at a smaller rate than the lowest EPs. The sequences of the lowest EPs for the two system sizes are much closer than the two sequences of the second lowest EPs. This pattern propagates to higher energies.

In Fig.~\ref{PADE} (e), we select a specific pair $(j_1, j_2)$ of eigenstates that coalesce and study $\alpha_{\text{EP}}(N, j_1, j_2)$ as a function of the system size. For large $N$, $\alpha_{\text{EP}}(N, j_1, j_2)$ changes almost continuously in the complex $\alpha$-plane. Using the Pad\'e extrapolation method, we can obtain numerically the limit of $\alpha_{\text{EP}}(N, j_1, j_2)$ for $1/N \rightarrow 0$. This method avoids the calculation of high order derivatives of $\alpha_{\text{EP}}$ with respect to $1/N$, as needed in Taylor and similar expansions~\cite{Schlessinger1968,Masjuan2013}. The limit for $\alpha_{\text{EP}}(N \rightarrow \infty, j_1, j_2)$ exists and equals the real value $\alpha_c=0.8$, as confirmed in Fig.~\ref{PADE} (e) for any of the chosen pairs. The convergence is faster for the EPs of lower energies. This shows that the critical point for the QPT can be obtained from non-Hermitian calculations considering finite system sizes.

As for the critical points of the ESQPT, we verified that the extrapolations of the vertical progressions of the EPs in Fig.~\ref{Fig:esptEP} (a) touch the curves of real eigenvalues (thin solid lines) and this happens very close to where these curves split. The line made of the intersection points between extrapolated EPs and real eigenvalues is nearly parallel to the separatrix and approaches it as the system size increases.

We detect the effects of EPs also in physical observables. For the real Hermitian Hamiltonian of finite systems, the behavior of quantities such as the total magnetization in the $z$ and in the $x$ direction changes abruptly, yet smoothly, close to the QPT and ESQPT critical values of the control parameter~\cite{Santos2016}. In the non-Hermitian approach, we find that by keeping $\rm{Im}(\alpha)$ fixed and varying $\rm{Re}(\alpha)$, a sudden non-analytical discontinuity in the values of those observables occur exactly when we reach the associated EP. Contrary to the Hermitian treatment, where non-analycities occur only in the thermodynamic limit, here they appear already for finite $N$. In finite system sizes, sharp non-analytical transitions associated with eigenvalues and eigenvectors can happen only in non-Hermitian quantum mechanics~\cite{MoiseyevBook}. In the thermodynamic limit, where the EPs fall into the real $\alpha$-axis, the results from the two approaches, Hermitian and non-Hermitian, coincide.

{\em Conclusions. --} Using a finite system described by the LMG model with the control parameter $\alpha$ continued into the complex plane, we showed that the EPs are linked with the avoided crossings that characterize the ground state QPT and ESQPTs obtained for the real (physical) LMG Hamiltonian.  These EPs approach the axis of real $\alpha$ in the thermodynamic limit. The eigenvalues pertaining to such EPs indicate the position of the separatrix that marks the ESQPT.

The approach presented here can be used for studying phase transitions in systems other than the LMG model. It should be of particular interest to models where the critical values are unknown and difficult to accurately determine from Hermitian methods.

\begin{acknowledgements}
This research was supported by the I-Core: the Israeli Excellence Center Circle of Light, by the Israel Science Foundation grants No. 298/11 and No. 1530/15, and by the American National Science Foundation grant No.~DMR-1147430 and No.~DMR-1603418. We thank Ofir Alon, Francisco P\'erez-Bernal, and Saar Rahav for fruitful discussions.
\end{acknowledgements}


\end{document}